\documentclass[structabstract]{aa} 
\usepackage{graphicx}
\usepackage{txfonts}
\usepackage{url}
\usepackage[round]{natbib}
\bibpunct{(}{)}{;}{a}{}{,}
\DeclareMathSymbol{\lesssim}{\mathrel}{AMSa}{"2E}
\begin{document}
\bibliographystyle{aa}
 \title{Polarimetric and spectroscopic optical observations of the ultra-compact X-ray binary 4U 0614+091 \thanks{Based on observations made with the Italian Telescopio Nazionale Galileo (TNG) operated on the island of La Palma by the Fundaci\'{o}n Galileo Galilei of the INAF (Istituto Nazionale di Astrofisica) at the Spanish Observatorio del Roque de los Muchachos of the Instituto de Astrofisica de Canarias and with the Nordic Optical Telescope, operated by the Nordic Optical Telescope Scientific Association at the Observatorio del Roque de los Muchachos, La Palma, Spain, of the Instituto de Astrofisica de Canarias and on observations made with ESO Telescopes at the Paranal Observatory under programme ID 079.D-0884(A).}}
   \author{M. C. Baglio 
          \inst{1, 2}          
          \and         
          D. Mainetti \inst{2, 3}
          \and
          P. D'Avanzo\inst{2}
          \and
          S. Campana \inst{2}
          \and
          S. Covino \inst{2}
          \and
          D. M. Russell \inst{4}
          \and
          T. Shahbaz \inst{5, 6}
          }
		  
   \institute{Universit\`{a} dell'Insubria, Dipartimento di Fisica, Via Valleggio 11, I–22100 Como, Italy                      \\
              \email{cristina.baglio@brera.inaf.it}
         \
             \\
             \and 
             INAF, Osservatorio Astronomico di Brera, Via E. Bianchi 46, I-23807 Merate (Lc), Italy
             \\
             \and
             Universit\`{a} di Milano-Bicocca, Dipartimento di Fisica G. Occhialini, Piazza della Scienza 3, I-20126 Milano, Italy
             \\
             \and
             New York University Abu Dhabi, P.O. Box 129188, Abu Dhabi, United Arab Emirates
             \\
             \and
             Instituto de Astrof\'{i}sica de Canarias (IAC), E-38200 La Laguna, Tenerife, Spain
             \\
             \and
             Departamento de Astrof\'{i}sica, Universidad de La Laguna (ULL), E-38206 La Laguna, Tenerife, Spain
\\             
              }
   \date{ }

   \abstract
   {}
   {We present a polarimetric and spectroscopic study of the persistent ultra compact X-ray binary 4U 0614+091 aimed at searching for the emission of a relativistic particle jet and at unveiling the orbital period $ P_{\rm orb} $ of the system. }
   {We obtained $ r $-band polarimetric observations with the Telescopio Nazionale Galileo (TNG) equipped with the PAOLO polarimeter and with the Nordic Optical Telescope (NOT) equipped with the ALFOSC instrument, covering $ \sim $2 hours and $ \sim $0.5 hours observations, respectively. We carried out low resolution spectroscopy of the system using the ESO Very Large Telescope equipped with FORS1 for $ \sim $1.5 hours (16 spectra covering the range 4300-8000 $ \AA $).}
   {The polarimetric analysis performed starting from the TNG dataset revealed a polarisation degree in the $ r $-band of $  3 \% \pm 1\% $. From the NOT dataset, due to the lower S/N ratio, we could obtain only a $ 3\sigma $ upper limit of 3.4$ \% $. From the joining of a spectroscopic and photometric analysis, through the study of the equivalent width variations of the CII 7240 $ \AA $ line and the $ r $-band light curve, we could find a hint of a $ \sim 45 $ min periodicity. }
   {A polarisation degree $ P $ of $ \sim 3 \% $ in the $ r $-band is consistent with the emission of a relativistic particle jet, which is supposed to emit intrinsically linearly polarised synchrotron radiation. Since no variations of $ P $ with time have been detected, and the accretion disc of the system does not contain ionised hydrogen, scattering by free electrons in the accretion disc has been rejected. The period of $ \sim 45 $ min obtained through the analysis of the system light curve and of the equivalent width variations of the selected spectral line is probably linked to the presence of a hot spot or a superhump in the accretion disc, and lead to an orbital period $ \gtrsim $ 1 hour for the binary system.
 }

\authorrunning{M. C. Baglio et al.} 
\titlerunning{Polarimetric and spectroscopic optical observations of 4U 0614+091}
\maketitle

\section{Introduction}\label{intro}

About thirty ultra-compact X-ray binaries (UCXBs) candidates have been identified to date \citep{Zand2007}. These systems are a subclass of low mass X-ray binaries (LMXBs) with very short orbital period, typically $\lesssim $ 80 min. The compact nature of UCXBs implies that the companion star's density is higher than typical main sequence stars, and so can be excluded (\citealt{Nelson1986}; \citealt{Savonije1986}). Because their sizes are comparable to the donor Roche lobe dimension, the most likely companion star candidates for UCXBs are white dwarfs or helium burning stars \citep{Nelson1986}. This inference was later confirmed through the observation of strong helium and carbon-oxygen lines in their spectra (\citealt{Schulz2001}; \citealt{Nelemans2004}; \citealt{Nelemans2006}).

The UCXB candidate 4U 0614+091 is a persistent LMXB \citep{Forman1978} that has been identified in the optical with a faint ($\sim18 \rm \, mag$), blue, variable star in the galactic plane \citep{Paradijs1994}. After the detection of Type I bursts \citep{Kuulkers2009}, the compact object of the system was identified as a neutron star. Assuming the Eddington luminosity for the bursts, a first estimate of the system distance was obtained ($< 3 \, \rm kpc$; \citealt{Brandt1992}), that was recently revised to 3.2 kpc \citep{Kuulkers2010}.
As expected for an UCXB, optical spectroscopy revealed significant carbon and oxygen lines, but no hydrogen or helium, leading to the conclusion that the system possesses a carbon-oxygen accretion disc (\citealt{Nelemans2004}; \citealt{Nelemans2006}). The observation of a broad emission line associated to O VIII Ly$\alpha$ emission in the X-ray band led to the identification of the companion star with an oxygen-rich donor star \citep{Nelemans2004}. However, the observation of type I bursts remained unexplained, since the presence of high quantities of helium is needed to account for these thermonuclear reactions. To solve this puzzle \citet{Juett2003} supposed that spallation reactions at the neutron star surface could split C and O nuclei into H and He which can then recombine again into heavier elements, explaining both the absence of H and He in the spectra and the occurrence of type I bursts.

Several attempts have been made to determine the orbital period of the system. Modulations in the optical/NIR band can be due to different phenomena, like the heating of one side of the companion by X-ray irradiation from the neutron star, but also to the presence of superhumps or hot spots from the impact point of the accretion stream with the disc. Firstly \citet{O'Brien2005} reported a $\sim$50 min orbital period from high-speed optical data obtained with ULTRACAM.  A $\sim$50 min period was also suggested from \citet{Zhang2012} thanks to the observation of a quasi-periodic oscillation. \citet{Shahbaz20084U} found evidence for three different periods based on optical photometry (42, 51.3 and 64 min) with the 51.3 min being the clearest modulation. 
 \citet{Hakala2011} was not able to find any periodicity for the system despite their extensive observations. \citet{Madej2013} finally noted a weak periodical signal modulated at a period of $ \sim$30 min in the red-wing/blue-wing flux ratio of the most prominent emission feature at $ \sim 4650\,\AA $. They proposed that this modulation could be due to Doppler shifts of the CIII and OII lines as well as variable flux ratio of the CIII with respect to the OII lines forming the feature. They hypothesized that this periodic signal could represent the orbital period of the source.

A recent work \citep{Migliari10} presented the most complete energy spectrum of 4U 0614+091, from radio to X-rays, suggesting the presence of optically thick and thin synchrotron emission from a jet of relativistic particles. Such a phenomenon is thought to be strictly bound to accretion \citep{Fender01} and in the LMXB scenario it is mostly expected to happen in persistent systems or in transient systems during outburst. The emission of a jet makes a polarimetric study of 4U 0614+091 particularly intriguing, since synchrotron emission is known to be intrinsically linearly polarised (\citealt{Ribicky79}).  

The paper is organised as follows: in Section 2 we present the results of the first optical polarimetric study (\textit{r}-band) performed on the persistent LMXB 4U 0614+091; in Section 3 we report our attempts to determine the orbital period of the system, through a photometric and spectroscopic analysis. All the errors are indicated at the $ 68 \% $ confidence level (c.l.), unless differently stated.

\section{TNG and NOT optical polarimetry}
The system 4U 0614+091 was observed on January 27, 2013 with the 3.6 m FGG TNG telescope at La Palma, equipped with the PAOLO polarimeter. A set of 28 images of 240 s integration each was taken, with the optical \textit{r} filter (6200 \AA). The night was clear, with seeing degrading with time (from $1.0''$ to $1.6''$). 4U 0614+091 was then observed a year later (March 11, 2014) in the same optical band with the ALFOSC instrument in polarimetric mode (using the Wedged Double Wollaston configuration, WeDoWo) mounted at the 2.5 m NOT telescope at La Palma, for a totality of 2 images of 900 seconds integration each.  
Image reduction was carried out following standard procedures: subtraction of an averaged bias frame, division by a normalized flat frame. Flux measurements have been performed through aperture photometry with \textit{\tt daophot} \citep{Stetson1987} for all the objects in the field.

Both the polarimeter PAOLO \citep{Covino_PAOLO} and the WeDoWo device \citep{Oliva1997} consist of a double Wollaston prism (DW) mounted in the filter wheel, which produces four simultaneous polarisation states of the field of view. The images are then separated by a special wedge, producing four image slices on the CCD that correspond to four different position angles with respect to the telescope axis (0$ ^{\circ} $, 45$ ^{\circ} $, 90$ ^{\circ} $ and 135$ ^{\circ} $). Such images possess all the information needed in order to provide linear polarisation measurements. This is a fundamental requirement for PAOLO, since the instrument is mounted on the Nasmyth focus of the TNG, introducing varying instrumental polarization of the order of 2-3 $ \% $ \citep{Giro2003}.

The normalised Stokes parameters for linear polarisation, $ Q $ and $ U $, are defined as follows:

\begin{equation}
Q=\frac{f(0^{\circ})-f(90^{\circ})}{f(0^{\circ})+f(90^{\circ})} ; \,\,\,\,\, U=\frac{f(45^{\circ})-f(135^{\circ})}{f(45^{\circ})+f(135^{\circ})},
\end{equation}
where $ f $ corresponds to the measured flux of the source.

An estimate of the observed linear polarisation degree of the incoming radiation can then be achieved from:
\begin{equation}\label{Peq}
P_{\rm obs}= \sqrt{Q^{2}+U^{2}},
\end{equation}
which should be corrected for a bias factor (\citealt{Wardle1974}; \citealt{Serego}) in order to account for the non-gaussianity of the statistic describing the polarisation distribution. In particular:
\begin{equation}\label{bias_corr}
P=P_{\rm obs}\sqrt{1-\left( \frac{\sigma_{\rm P}}{P_{\rm obs}}\right)^{2} },
\end{equation}
where $ \sigma_{\rm P} $ is the r.m.s. error on the polarisation degree.

The polarisation angle $ \theta $ can be obtained from the relation:
\begin{equation}\label{pol_angle}
\theta = 0.5 \tan^{-1}(U/Q).
\end{equation}
\subsection{Average degree of linear polarisation}\label{ave_degree_sec}

We considered first of all the set of images obtained in 2013 with the PAOLO instrument. In order to reach a high S/N ratio, that is fundamental in polarimetry, we tried to estimate the linear polarisation degree in the $ r $-band starting from the average of all the images with good seeing. In particular, we defined a limit to the seeing of $ \lesssim 1.3'' $, which allowed us to use 12 images (48 min). 

\begin{figure}[!h]
\begin{center}
\includegraphics[scale=0.27]{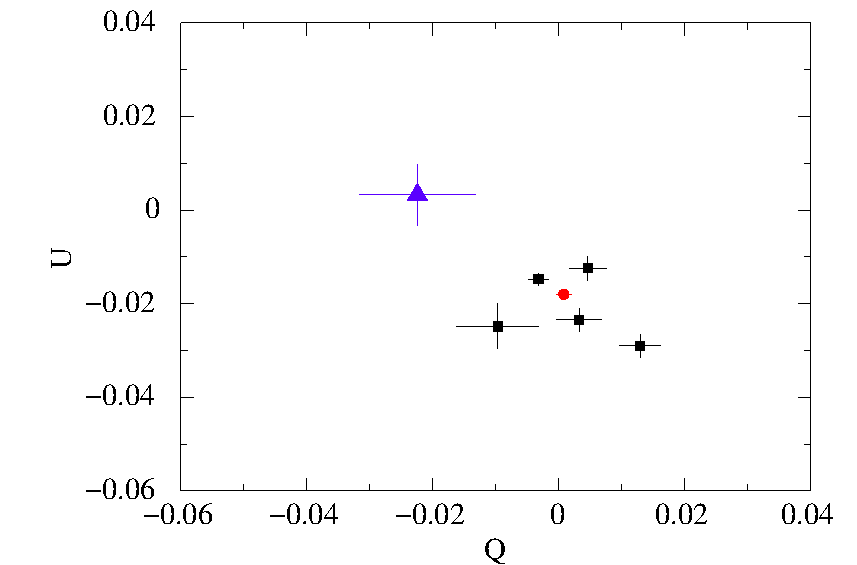}
\caption{$ U $ vs. $ Q $ for the averaged image of the optical $ r $ filter, for 4U0614+091 (blue triangle) and for five reference field stars (black squares), Tab. \ref{tab_Q_U}. With a red dot we indicated the weighted mean of the reference stars Stokes parameters. The parameters reported are not corrected for interstellar or instrumental effects.}
\label{U_Q_media}
\end{center}
\end{figure}

The Stokes parameters $ Q $ and $ U $ of all the objects in the field have been evaluated by means of the instrumental polarization model extensively described in \citet{Covino_PAOLO}. Since the field stars we chose as references cluster rather well around a common value in the $ Q-U $ plane (Fig. \ref{U_Q_media}), we could safely assume them to be intrinsecally unpolarised and that the interstellar polarisation for these bright stars in the field is probably low. This hypothesis is supported from the evaluation of the instrumental contribution only to the polarisation of the field stars, that was possible thanks to the tools described in \citet{Covino_PAOLO} and that was consistent with the results obtained for the Stokes parameters of the chosen reference stars. In particular, the modelling of the instrumental polarization by means of observations of a suitable set of polarimetric (polarized and/or unpolarized) standard stars has shown routinely a rms residual at about 0.2\% or even better if the observations cover a rather limited range in Hour Angles, as this is the case.
This fact allowed us to correct the values of $ Q $ and $ U $ of the target for the average $ Q $ and $ U $ obtained for the field stars, in order to account for the not negligible effects of instrumental polarisation. The amount of this correction is reported in Tab. \ref{tab_Q_U} as the weighted mean.

\begin{table}
\caption{Values of the Q and U Stokes parameters not corrected for interstellar or instrumental effects represented in Fig. \ref{U_Q_media} for 4U0614+091 and for five reference field stars. The weighted mean of the reference stars Stokes parameters has been reported in the last raw and corresponds to the amount of correction that we applied to $ Q $ and $ U $ of the target.} 
\label{tab_Q_U}
\centering                          
\begin{tabular}{|c c c|} 
\hline\hline
Object & Q ($ \% $) & U ($ \% $)  \\
\hline
4U0614+091 & $-2.24 \pm 0.93$ & $0.32 \pm 0.66$ \\
Star 1 & $-0.32 \pm 0.17$ & $-1.47 \pm 0.13$\\
Star 2 & $0.46 \pm 0.29$ &$-1.24 \pm 0.26$ \\
Star 3 & $1.29 \pm 0.33$ & $-2.90 \pm 0.26$\\
Star 4 & $-0.97 \pm 0.65$ & $-2.50 \pm 0.49$\\
Star 5 & $0.32 \pm 0.36$ & $-2.35 \pm 0.26$\\
Weighted mean & $0.08 \pm 0.12$ & $-1.80 \pm 0.10$\\
\hline 
\end{tabular}
\end{table}

Starting from these Stokes parameters (that can be supposed to be normally distributed), we used a Monte Carlo simulation to obtain the probability distribution that describes the polarisation degree $ P $ of the radiation (Rice distribution, \citealt{Wardle1974}). Passing from a Cartesian coordinate system with axis \textit{Q} and \textit{U} to a polar system, where $ P $ is the radius (eq. \ref{Peq}), we had to correct the distribution for the geometrical factor 1/$ P$, obtained evaluating the Jacobian determinant of the coordinate change. The probability density function obtained after this correction can be demonstrated to be at first approximation a Gaussian, simply supposing the errors of \textit{Q} and \textit{U} to be similar between each others in the Rice distribution.
From the fit of this new distribution with a Gaussian function, we evaluated the most probable value of $ P $ and its uncertainty to be $ P=2.85 \pm 0.96 \%$ that is significant at a $ 3\sigma $ level. This value does not need any bias correction (eq. \ref{bias_corr}), since it has been evaluated without supposing $ P $ to be normally distributed.
Furthermore we obtained an estimate of the polarisation angle $ \theta $ (eq. \ref{pol_angle}) of $ 85.9^{\circ} \pm 8.4^{\circ} $.
Following \citet{Serkowski75} we could evaluate the maximum expected interstellar contribution to the linear polarisation $ P_{\rm max} $ of 4U 0614+091. In particular we made use of the empirical formula $ P_{\rm max} \leq 3A_{\rm V}$, where $ A_{\rm V}=1.4 $ is the total $ V $-band Galactic extinction expected along the source line of sight\footnote{\url{http://ned.ipac.caltech.edu/forms/calculator.html}}. According to this relation, the maximum contribution to the linear polarisation for 4U 0614+091 due to interstellar effects should remain under $ 4\% $. For this reason it is not possible to fully rule out the possible involvement of the interstellar dust in the observed polarisation degree of the target. 
Future multi-wavelength observations will allow us to obtain a polarimetric spectral energy distribution, that, if fitted with the Serkowski law \citep{Serkowski75}, will permit us to verify whether the observed polarisation degree has an interstellar origin or is intrinsic to the target.


We then performed a similar analysis on the two images taken with the WeDoWo device in 2014, summing them together in order to achieve a higher significance. The Stokes parameters have been extracted thanks to another custom set of command-line tools developed for the WeDoWo instrument, and were corrected through the observation of the standard non-polarised star G191B2B\footnote{\url{http://www.not.iac.es/instruments/turpol/std/zpstd.html}}. The polarisation degree $ P $ was obtained through the same analysis described above. Unfortunately a combination between the faintness of the source and the bad seeing of the night ($ >1.5'' $) did not permit us to obtain a significant polarisation detection for the source. A $ 3\sigma $ upper limit to $ P $ of the 3.4$ \% $ was obtained, that is consistent with the result achieved for the 2013 dataset.

 
\subsection{Time-dependent linear polarisation}\label{pol_light curve_sec}
\begin{figure}
\begin{center}
\includegraphics[scale=0.4]{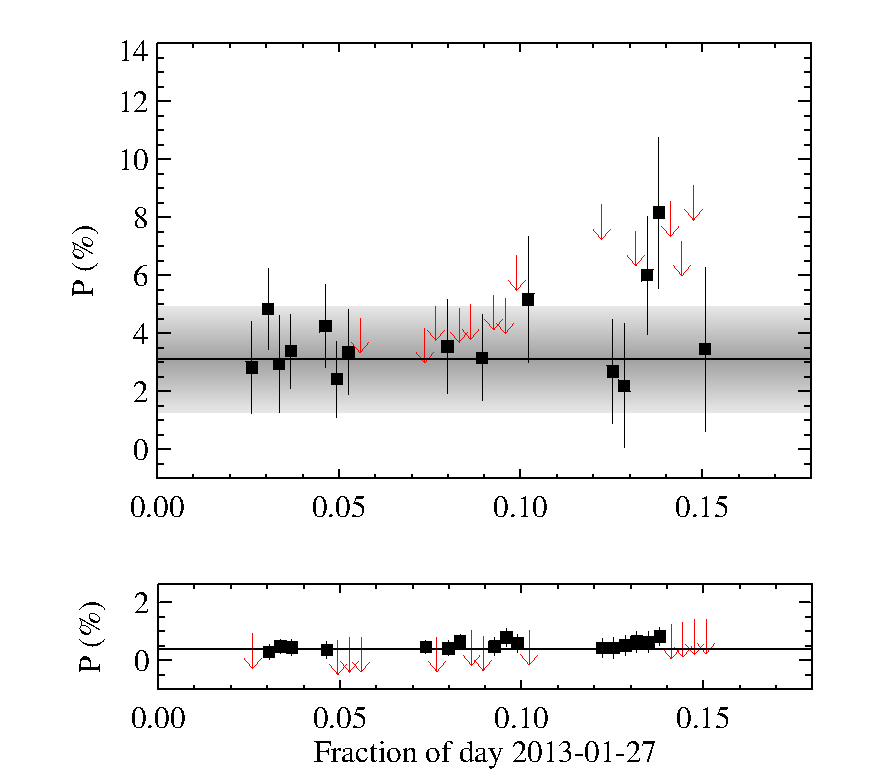}
\caption{\textit{Top} panel: $ r $-band polarisation curve of 4U0614+091. With black squares we reported the polarisation detection ($ P/\sigma_{\rm P}\geqslant 1 $) with their error bars, whereas with red arrows we represented all the upper limits that we evaluated when a polarisation measure was not possible, due for example to lower S/N ratios. The superimposed horizontal line represents the value $ P_{\rm mean} $ obtained by fitting the polarisation detections and upper limits with a constant. The gray band represents the 1$ \sigma $ level. Note that a better determination of the polarisation level is obtained from the probability density function (as discussed in the text), resulting in a similar central value but a much smaller error ($0.96\%$). \textit{Bottom} panel: $ r $-band polarisation curve of one field star chosen as reference (black squares: polarisation detections; red arrows: upper limits).}
\label{upper}
\end{center}
\end{figure}

With the aim of detecting some kind of variability in the polarisation degree of the source, we analised the 28 images obtained with the PAOLO polarimeter, extracting the normalised Stokes parameters $ Q $ and $ U $ as above. The polarisation degree trend as function of time for the target and for one of the reference stars is shown in Fig. \ref{upper}. For $ \sim $ the $ 50 \% $ of the images the target did not possess a polarisation degree different from 0 at a $ \geqslant \, 1 \sigma $ level; in that cases we decided to report in Fig. \ref{upper} the $ 3\sigma $ upper limits on $ P $. 

The polarisation degree of 4U 0614+091 does not show any particular trend with time, remaining almost constant around a common value $ P_{\rm mean} $. In order to take into account both the polarisation detections and upper limits in Fig. \ref{upper} (\textit{top} panel), and maintaining the same cut to seeing $ \lesssim 1.3 $ used in Sec. \ref{ave_degree_sec}, we obtained $ P_{\rm mean} $ by summing together the distributions of polarisation of the selected images and by fitting the obtained distribution with a Gaussian function. With this method, $ P_{\rm mean}= 3.1 \% \pm 1.8 \%$, that is consistent with the value of $ P $ obtained in Sec. \ref{ave_degree_sec}.

\section{The orbital period of 4U 0614+091}

\subsection{$ r $-band light curve of 4U0614+091}
As stated in Sec. \ref{intro}, no decisive measure of the orbital period $ P_{\rm orb} $ of 4U 0614+091 has been obtained to date. A possibility for measuring $ P_{\rm orb} $ is to observe a periodic modulation in the optical flux emitted from the source. In fact in the case of LMXBs, where the dominant optical/IR emission from a system in quiescence is the companion star, they are often subject to ellipsoidal modulations. This effect derives from the fact that the companion star suffers for a tidal distortion due to the large gravitational field of the compact object, and for this reason the projected surface area of the distorted star is different at quadratures than at conjunctions, resulting in maxima and minima of the observed flux, respectively. When the systems possess shorter orbital periods and smaller orbital separations (as in case of UCXBs), supposing the companion star emission to be the most relevant component, the effects of irradiation should dominate the observed fluxes, causing the light curves to have a single maximum and a single minimum around the phases of superior and inferior conjunction, respectively. Observing these patterns would allow us to measure precisely the orbital period of the source. However, such methods are usually effective with quiescent LMXBs, where the companion star is expected to dominate the optical emission.

The photo-polarimetric images of 4U 0614+091 taken with PAOLO allowed us to compute the system $ r $-band light curve. Infact, not considering a negligible loss in flux, the sum of the intensities measured in the four slices for each image results in the total flux coming from our target of observation. 
We therefore summed the fluxes corresponding to the four different position angles for the system 4U 0614+091 and for 5 isolated stars in the field of view. We performed differential photometry with respect to this selection of reference stars, in order to minimize any systematic effect. 
The calibration of the magnitudes was performed using the R2 magnitudes of the USNO B1.0 catalogue\footnote{\url{http://www.nofs.navy.mil/data/fchpix/}}, whose magnitudes were transformed into $ r $-band magnitudes using the transformation equations of \citet{Jordi2006}. However, a systematic error of a few tenths of a magnitude should be taken into account due to the inaccuracy of the USNO magnitudes.

\begin{figure}
\begin{center}
\includegraphics[scale=0.42]{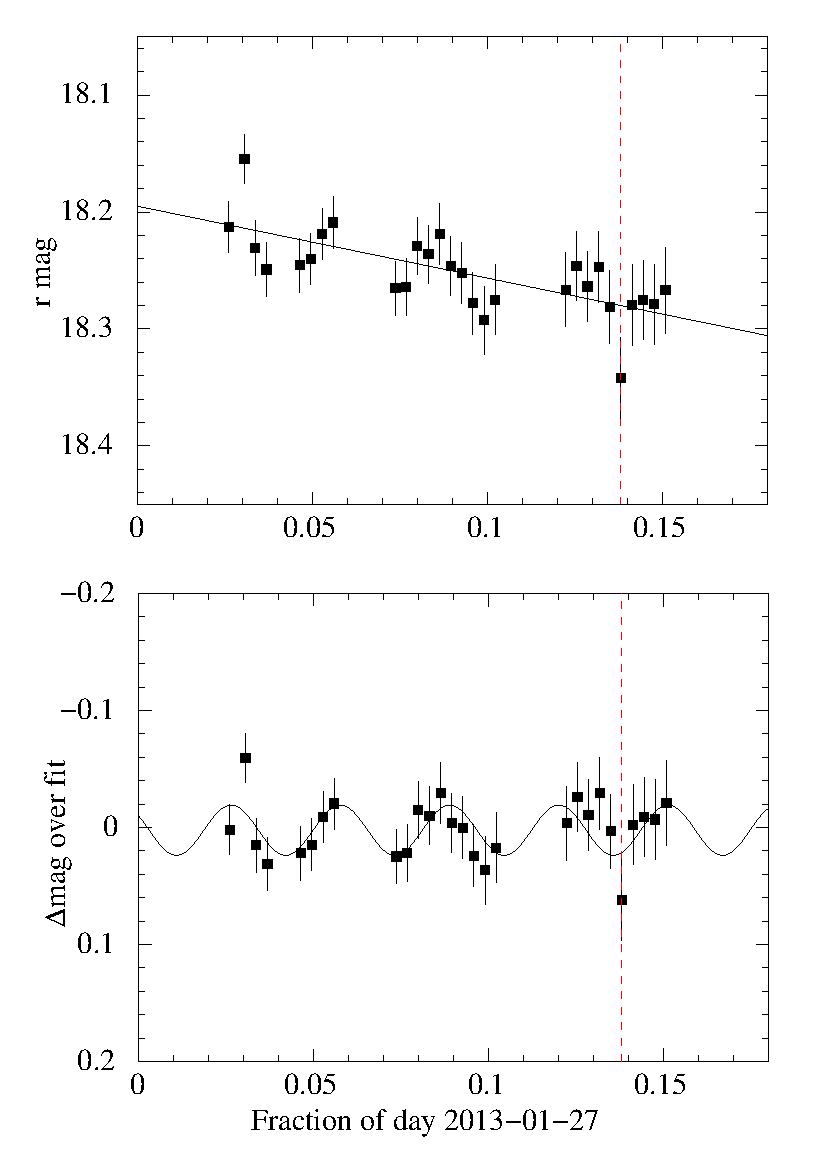}
\caption{\textit{Top} panel: $ r $-band light curve obtained for the system 4U0614+091. Superimposed, the fit of the dataset with a straight line. \textit{Bottom} panel: residual light curve after the subtraction of the straight line obtained with the fit of the light curve in the \textit{top} panel from the dataset itself. Superimposed, the fit of the curve with a sinusoidal function + a constant. With a dashed line we indicated in both panels the fraction of the day that corresponds to the possible polarisation flare (Fig. \ref{upper}.}
\label{light curve}
\end{center}
\end{figure}

The light curve (Fig. \ref{light curve}, \textit{top} panel) shows a general decreasing trend with time, probably due to accretion variations in the disc, with superimposed a significant short term variability. The fit with a straight line $y=mx+q$ produces a $ reduced $ $ \chi ^{2}$ of $ \sim 1.04 $, with $ q= 18.19 \pm 0.01$ and $ m=0.61\pm 0.13 $. We could then subtract from the light curve the straight line obtained from the fit, ending up with the residuals in Fig. \ref{light curve} (\textit{bottom} panel) that suggest a sinusoidal modulation (with significance probability of $ \sim 98\% $ given by an f-test) with a periodicity of $ 44.9 \pm 2.3 $ min and a semi-amplitude of $ (2.15\pm 0.73) \times10^{-2} $ mag (reduced $ \chi ^{2}$ of $ \sim 0.67 $).
 

\subsection{Spectroscopy with FORS1}\label{spectroscopy}
\begin{figure}[!h]
\centering
\includegraphics[scale=0.42]{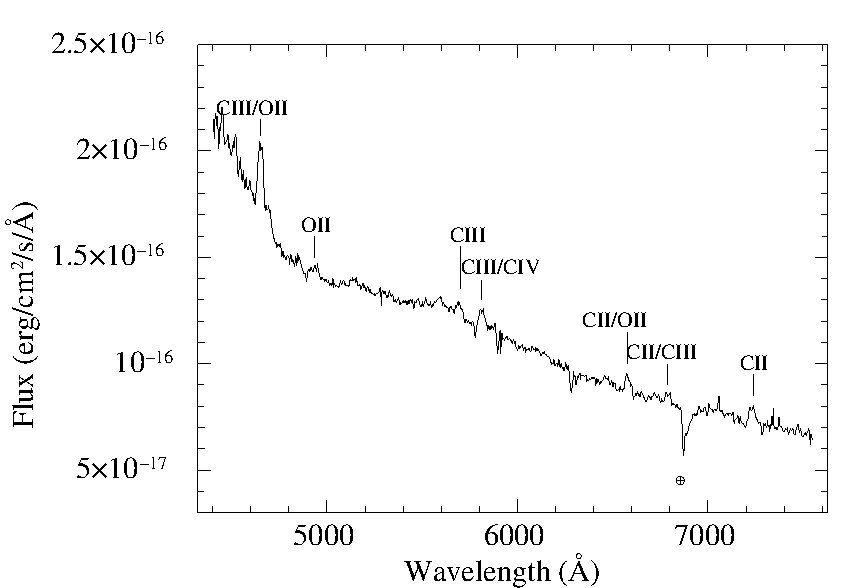}
\caption{Average spectrum of 4U 0614+091 with indicated the most prominent identified emission lines (Tab. \ref{tab_spectrum}).}
\label{spectrum}
\end{figure}

A set of 16 low resolution ($ \sim 160\,\, \rm km\, s^{-1} $) spectra of 300 s integration each was taken of 4U 0614+091 on 3 September 2007 with the ESO VLT (Very Large Telescope), equipped with the FORS1 spectrograph. All the spectra were taken with the 300V grism with a 1-arcsec slit, using 2 $ \times $ 2 on-chip binning and covering the wavelength range 4300-8000 \AA. The night was clear, with seeing variable around a mean value of $ \sim 1.2 '' $. These observations cover $ \sim 1.5 $ times the possible $ \sim 50 $ minutes orbital period of the system. The extraction of the spectrum was performed with the ESO-MIDAS\footnote{\url{http://www.eso.org/projects/esomidas/}} software
package. Wavelength and flux calibration of the spectra were achieved using helium-argon lamp and observing spectrophotometric stars.
In Fig. \ref{spectrum} we report the average spectrum of the source. The spectrum shows both absorption and emission lines, the latter possibly due to the presence of partially ionised carbon and oxygen, that possess a lot of lines in the investigated wavelength range.

We could not find any trace of hydrogen and helium lines in our average spectrum, leading to the conclusion that the companion star of the system must be a hydrogen-poor star. The identification of the emission lines was made by the comparison with the accurate results of \citet{Nelemans2004}, that in turn used the local thermodynamic equilibrium (LTE) model proposed by \citet{Marsh1991}. In particular, our results are in agreement with the companion star being a C/O-rich white dwarf as stated in \citet{Nelemans2004}. The most prominent identified features in our spectrum are reported in Tab. \ref{tab_spectrum}.

\begin{table}
\caption{Strongest features identified in the spectrum of 4U 0614+091 (Fig. \ref{spectrum}). For details about the transitions, see Tab. 3 in the work of \citet{Nelemans2004}.} 
\label{tab_spectrum}
\centering                          
\begin{tabular}{|c c|} 
\hline\hline
Feature (\AA) & Ion \\
\hline
4650  & CIII\\
      & OII \\
4700  & OII \\
4935  & OII \\
5700  & CIII\\
5810  & CIII\\
      & CIV?\\
6580  & CII \\
      &(OII)\\
6790  & CII \\   
7240  & CII \\

\hline 
\end{tabular}
\end{table}

\begin{figure}[!h]
\centering
\includegraphics[scale=0.3]{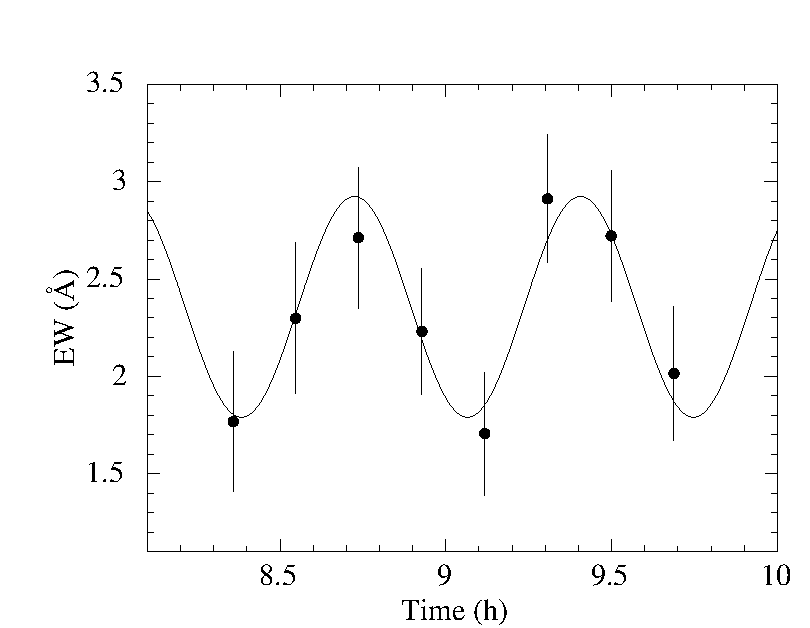}
\caption{EW variation curve of CII 7240 $\AA$ line, obtained by averaging the spectra two by two, with superimposed a sinusoidal fit with period 41 min.}
\label{EWfit}
\end{figure}

In order to determine the orbital period of the source, we studied the variations of the equivalent width (EW) of the emission lines CII/OII 6580 \AA $\,$ and CII 7240 \AA, indicated in Fig. \ref{spectrum}. We performed sinusoidal fitting of the EW variations, averaging the spectra two by two with the aim of enhancing the signal to noise ratio. The results of the fits are reported in Tab. \ref{tab_fit_sin}: the most significant periodicity is the obtained with the 7240 \AA $\,$line (Fig. \ref{EWfit}), for which $ P=40.9 \pm 6.8 $ min, that is consistent with the period of the oscillation obtained from the $ r $-band light curve (Fig. \ref{light curve}, \textit{bottom} panel). We also tried to measure the radial velocity of the CII/OII 6580 $\AA$ and CII 7240 $\AA$ emission lines. We found no evidence for periodical Doppler motions, likely due to the poor resolution of our spectra.
\begin{table}
\caption{Results of the sinusoidal fit of the EW variations for the two selected lines. In the last column we reported the confidence level of the fit, obtained with an F-test with respect to a fit with a constant. All the uncertainties are reported at the 90$ \% $ confidence level. } 
\label{tab_fit_sin}
\centering                          
\begin{tabular}{|c c c c|} 
\hline\hline
$\lambda$ (\AA) & Period (min) & Reduced $ \chi ^{2} $ & CL \\
\hline
6580 & $37.9 \pm 2.6$ & 5.98 & 45.7 $\%$\\
7240 & $40.9 \pm 6.8$ & 0.29 & 1.5 $ \% $\\
\hline 
\end{tabular}
\end{table}

\section{Discussion}
\subsection{Polarimetric signatures of a jet}
Relativistic particle jets have been observed to be emitted from different kinds of systems, like Active Galactic Nuclei (AGN), Super Soft Sources and X-ray binaries (XRBs), all subject to the phenomenon of accretion (disc-jet coupling, \citealt{Fender01}).
The existence of jets in XRBs in particular does not seem to depend on the type of compact object hosted in the system, that can be both a neutron star and a black hole (recently evidence for a transient jet in a binary
system containing a white dwarf has been found, \citealt{Kording08}).
However, XRBs hosting black holes are among the most studied jet emitters sources, also because they are brighter then, e.g, neutron stars systems in the wavelength range where the jet contributes the most to the emission, i.e. the radio band. Generally, XRBs that emit jets are characterised by an optically thick, flat synchrotron radio spectrum, interpreted as a signature of the presence of the compact jet (for some galactic black hole XRBs this has been confirmed by radio imaging of the jet itself; \citealt{Stirling01}; \citealt{Dhawan2000}). Because the size scale of the emitting region in the jet is expected to scale inversely with frequency (\citealt{Nowak2005}; \citealt{Blandford79}), the jet break frequency marks the start of the particle acceleration in the jet \citep{Polko2010}.
A break from an optically thick to optically thin synchrotron spectrum is then expected in the IR/optical range \citep{Falcke2004}.
The break is detected in 4U0614+09 \citep{Migliari10} and also seen in the black hole candidates GX 339--4 (\citealt{Corbel02}; \citealt{Gandhi11}; \citealt{Corbel2013}), XTE J1118+480 \citep{Hynes2006}, XTE J1550-564 \citep{Chaty2011}, V404 Cyg \citep{Russell2013a} and MAXI J1836-194 \citep{Russell2013b}.
 

Synchrotron emission is expected to be intrinsically linearly polarised at a high level, up to tens of per cent, especially in case of ordered magnetic fields. A polarimetric signature of synchrotron-emitting jets can be observed both in the NIR and in the optical in LMXBs; however, due to tangled and turbulent magnetic fields at the base of the jets, polarisation degrees in these sources have never been found to exceed a few per cent, as stated in \citealt{Russell11}. The only evidence of ordered magnetic fields in LMXBs jets has been observed for the persistent system Cyg X-1 \citep{Russell14}. Only a few LMXBs have been observed with polarimetric techniques to date, both persistent and transient ones (\citealt{Charles80}; \citealt{Dolan89}; \citealt{Gliozzi98}; \citealt{Hannikainen00}; \citealt{Schultz04}; \citealt{Brocksopp07}; \citealt{Shahbaz08}; \citealt{Russell08}; \citealt{Russell11}; \citealt{Baglio_cen2014}). 

The aim of our work was to obtain additional evidence for the emission of a compact jet from the UCXB 4U 0614+091. Since the presence of a jet was stated from \citet{Migliari10}, we expected to observe a degree of linear polarisation of at least a few $ \% $, according to previous observations of polarised XRBs, due to the jet synchrotron radiation for a tangled magnetic field. With our analysis, we measured an average polarisation degree of $ 2.85 \% \pm 0.96 \% $, as expected. 

We should also consider that in systems where an accretion disc is present, not only synchrotron radiation from a relativistic particles jet could be the cause of emission of polarised light. In particular, hydrogen in the accretion disc is expected to be almost totally ionised, due to the high temperatures caused by viscosity. In this state, Thomson scattering of the emitted radiation with these free electrons is expected, causing a polarisation degree of at most a few per cent in the optical (\citealt{Brown78}; \citealt{Dolan84}; \citealt{Cheng88}). However, no hydrogen is present in the accretion disc of 4U 0614+091, since the companion star of the system has been found to be an oxygen-rich mass donor star, and hydrogen and helium lines have never been observed.

We deduce that the significant non-zero degree of polarisation of 4U0614+091 can be due to synchrotron radiation from a relativistic particle jet emitted from its central regions, which is inferred from the SED of the system \citep{Migliari10}. 

\begin{figure}[!h]
\centering
\includegraphics[scale=0.38]{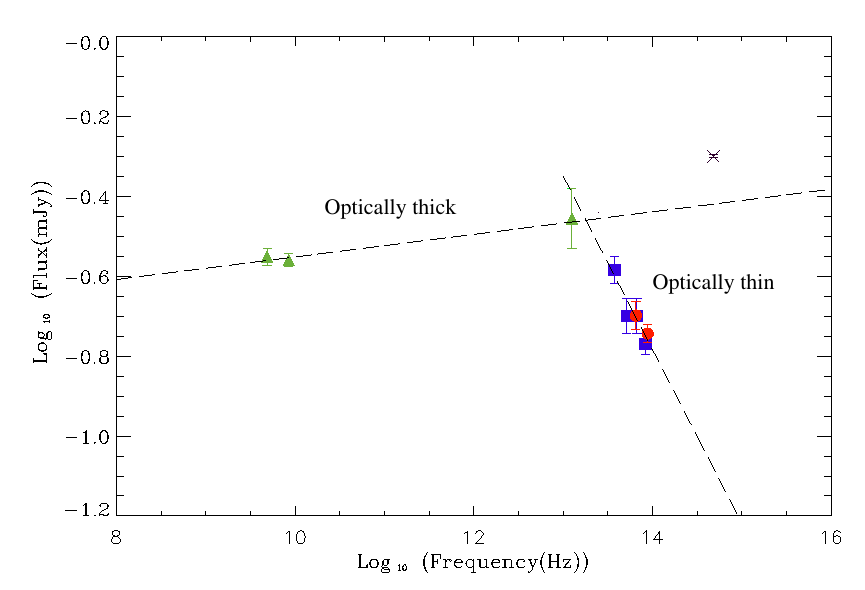}
\caption{Spectral energy distribution of the system 4U 0614+091 built starting from the radio (green triangles) and infrared points (blue squares) of \citet{Migliari10}, from the ALL WISE catalog magnitudes (red circles) and from the $ r $-flux obtained in this work (black x). With an orange arrow we indicated the W3-band WISE upper limit obtained from the ALLWise catalog. Superimposed, the two linear fits of the SED. All the points are normalized to the fluxes of \citet{Migliari10}.}
\label{sed}
\end{figure}

Following \citet{Migliari10}, we built the infrared SED of 4U 0614+091 adding to the infrared and radio points reported in that paper and the archival data in the ALLWISE catalog (Tab. \ref{tab_sed}; Fig. \ref{sed}). These data are not contemporary to that of \citet{Migliari10}, but they fit well with the previous dataset, meaning that the average spectrum of the target does not vary significantly with time. From the fit with a linear function we obtained a spectral index $ 	\alpha $ of $ \sim $0.03 in the optically thick part of the spectrum, whereas in the optically thin part $ \alpha  \sim -0.43$, consistent with \cite{Migliari10}. The break frequency $ \nu_{\rm break} $ between optically thin and optically thick synchrotron emission remains in the same range of \citealt{Migliari10} ($ 1.25\times10^{13} \rm Hz < \nu_{\rm break}< 3.71\times10^{13} \rm Hz$).

\begin{table}
\caption{RADIO and IR fluxes (uncorrected for the negligible Galactic reddening)} obtained from the ALLWISE catalogue and from \citet{Migliari10} and $ r $-band de-reddened flux ($ A_{\rm V}=1.4 $) from the TNG data used to build the SED in Fig. \ref{sed}. 
\label{tab_sed}
\centering                          
\begin{tabular}{|c c |} 
\hline\hline
Band & Flux (mJy)  \\
\hline
4.86 GHz \citep{Migliari10} & $0.281 \pm 0.014$ \\
8.46 GHz \citep{Migliari10} & $0.276 \pm 0.010$ \\
\hline
$1.250 \times 10^{13}$ Hz (24 $\mu m$, \citealt{Migliari10}) & $0.35 \pm 0.06$ \\
\hline
$3.75 \times 10^{13}$ Hz (8 $\mu m$, \citealt{Migliari10}) & $0.26 \pm 0.02$ \\
$5.17 \times 10^{13}$ Hz (5.8 $\mu m$, \citealt{Migliari10}) & $0.20 \pm 0.02$ \\
$6.67 \times 10^{13}$ Hz (4.5 $\mu m$, \citealt{Migliari10})  & $0.20 \pm 0.02$ \\
$8.33 \times 10^{13}$ Hz (3.6 $\mu m$, \citealt{Migliari10})  & $0.17 \pm 0.01$ \\
\hline
$8.8 \times 10^{13}$ Hz (ALLWISE W1) & $0.18 \pm 0.01$ \\
$6.5 \times 10^{13}$ Hz (ALLWISE W2) & $0.20 \pm 0.02$ \\
$2.5 \times 10^{13}$ Hz (ALLWISE W3) & $<$ 0.36\\
\hline
$4.82 \times 10^{14}$ Hz ($ r $-band, this work)& $0.49 \pm 0.01$\\
\hline 
\end{tabular}
\end{table}

Under the hypothesis of jet emission, starting from the linear fit to the infrared part of the SED in Fig. \ref{sed} we could estimate the expected $ r $-band flux due to the jet emission ($F_{\rm jet}\sim 0.1$ mJy). Since we know that the total de-reddened $ r $-band flux of the source at the time of the TNG polarimetric observation was $ F_{\rm r}\sim $ 0.5 mJy, we could obtain from the ratio between $F_{\rm jet}$ and $ F_{\rm r}$ an estimate of the intrinsic linear polarisation degree of the jet only of $\sim 15 \% $ .

The polarisation light curve (Fig. \ref{upper}, \textit{top} panel) shows an almost constant trend with time, as stated in Sec. \ref{pol_light curve_sec}. Nevertheless it has to be noted that for higher times a hint of a small polarisation flare seems to be present; if we suppose the emission of the jet to be constant with time, the decrease of the flux that is observed in Fig. \ref{light curve} contemporary to the possible polarisation flare (indicated with a dashed line) makes the percentile flux of the jet increase, and this could explain the possible higher polarisation degree observed.

\subsection{Attempts in determining 4U0614+091 orbital period}\label{attempts}

\citet{Shahbaz20084U} found evidence of three different orbital periods for 4U0614+091, with the clearest modulation at 51.3 minutes likely due to a superhump rather than the orbital period of the system. With the aim of eliminating any ambiguity, we tried to obtain an estimate of the system orbital period by means of photometric and spectroscopic observations. We obtained the $ r $-band light curve of the system starting from the polarimetric measurements, and we could observe a slight decrease of the flux with time (Fig. \ref{light curve}, \textit{top} panel), with a possible superimposed sinusoidal modulation with a periodicity of $ \sim 45 $ min (Fig. \ref{light curve}, \textit{bottom} panel). This modulation could arise from the X-ray irradiation of the inner face of the companion star from the compact object (in this case, giving direct information about the orbital period of the binary) or from the presence of hot spots or superhumps in the accretion disc, that can create a modulation in the light curve due to the rotation of the system. 

We then tried to identify any periodicity in the system through the analysis of the EW variation curves of the CII 7240 $\AA$ line. From a sinusoidal fit (Fig. \ref{EWfit}) we obtained a periodical modulation at a $40.9 \pm 6.8$ minutes period, that corresponds to the orbital period of the line-emitting outer regions of the disc, and is consistent with the one measured from the optical light curve (Fig. \ref{light curve}, \textit{bottom} panel). This suggests that both the modulations could be caused by the same phenomenon; specifically, we propose that the EW variation of the CII 7240 $\AA$ emission line could be due to the modulation of the continuum emission, that we observe indeed in our light curve, and that should be linked to the accretion disc, as our spectroscopic analysis pointed out. Following this hypothesis, we could rule out the X-ray irradiation as the origin of the observed modulation in the light curve, that can be thus explained e.g. referring to the presence of hot spots or superhumps in the accretion disc.
Unfortunately the calibration precision of our spectra is too low in order to detect a modulation in the continuum emission as small as the one that we observed in the light curve (i.e. $ \sim 2.15\times10^{-2} $ mag), whose amplitude of oscillation does not exceed $ 10\% $.

Since the periodicity that we measured from the EW variation is linked to the outer regions of the disc, in order to obtain an estimate of the orbital period of the source we approximated the outer radius of the disc with its tidal radius $ R_{\rm tid} $ \citep{King96}:
\begin{equation}
R_{\rm tid}=0.9R_{\rm L},
\end{equation}
where $ R_{\rm L} $ is the Roche lobe radius of the compact object of the system, and is given by:
\begin{equation}\label{roche_lobe_eq}
\frac{R_{\rm L}}{a}=\frac{0.49\, q^{2/3}}{0.6\, q^{2/3} + \ln \left(1+q^{1/3}\right)},
\end{equation}
where $ q=M_{\rm NS}/M_{*} $ is the ratio between the neutron star and the companion star mass ($ M_{\rm NS} $ and $M_{*}$, respectively) and $ a $ is the binary separation, that is an upper limit to the radius of the companion star orbit.

From the Kepler law, we know that:
\begin{equation}\label{orb_period}
\frac{P(R_{\rm tid})}{P_{\rm orb}}\propto \left( \frac{R_{\rm out}}{a}\right)^{3/2} ,
\end{equation}
where $ P(r_{\rm tid}) $ and $ P_{\rm orb} $ are, respectively, the orbital period of the outer region of the disc and of the companion star of the system. 
Considering a mass ratio of at most $\sim 0.1 $, that is typical of a UCXB system, we obtain $ P_{\rm orb} \gtrsim 2P(r_{\rm tid})$, depending on the value of $ q $ that we consider, and meaning that the orbital period of our system should exceed 1 hour.

\section{Conclusions}
In this work, we presented the results of a $ r $-band polarimetric and a spectroscopic analysis of the persistent UCXB 4U0614+091, based on observations obtained in 2007, 2013 and 2014 with the VLT FORS1 spectrograph, the TNG PAOLO and the NOT WeDoWo polarimeters, respectively. 

In \citet{Migliari10} the authors stated the presence of an infrared excess in the spectral energy distribution of the system, that they interpreted as the signature of synchrotron emission from a relativistic particles jet. We were for this reason interested in searching for an optical linear polarisation degree of the order of some per cent, since synchrotron emission is expected to be intrinsically linearly polarised at this level. We obtained a $ r $-band polarisation degree of $ 2.85 \% \pm 0.96 \% $ from the TNG dataset, and a 3$ \sigma $ upper limit of 3.4$ \% $ from the NOT data. A polarisation degree of a few per cent in the optical is exactly what is expected in case of a jet emission; for this reason we can confirm the presence of this further component in the system emitted radiation. Under this hypothesis, we could moreover estimate an intrinsic linear polarisation degree of the jet only in the $ r $-band of $ \sim 15\% $.

We then tried to determine the system orbital period. We built the light curve of 4U 0614+091 starting from the polarimetric images obtained with the PAOLO polarimeter, observing a decreasing trend of the flux with time, possibly due to accretion variations in the disc, with superimposed a sinusoidal modulation at a $ \sim 45 $ min period. From the spectroscopic FORS1 images we could then extract the EW variation curve of the CII 7240 $ \AA $ line, and from its sinusoidal fit we could obtain a periodicity of $ 40.9 \pm 6.8 $ min, that refers to the outer regions of the disc and is consistent with the period of the modulation obtained with the optical light curve, suggesting for a common origin of the two periodicities (probably a hot spot or a superhump in the accretion disc). From the Kepler law, using the tidal radius of the accretion disc as an approximation for its outer radius, we could estimate for the system 4U0614+091 an orbital period $ \gtrsim $ 1 hour. 

\begin{acknowledgements}
MCB acknowledges S. Crespi for helpful discussions and T. Pursimo and I. Andreoni for their support during her observing run in La Palma. DM acknowledges Prof. M. Colpi (Universit\`{a} di Milano-Bicocca) for supportive discussions and the INAF-Osservatorio Astronomico di Brera for kind hospitality during her bachelor thesis. TS was supported by the Spanish Ministry of Economy and Competitiveness (MINECO) under the grant  AYA2010-18080.
\end{acknowledgements}

\addcontentsline{toc}{chapter}{Bibliografia}


\end{document}